\documentclass[prl,twocolumn,showpacs,amsmath,amssymb]{revtex4}
\usepackage{latexsym}
\usepackage[dvips]{graphicx}
\usepackage{dcolumn}
\usepackage{bm}

\begin{document}

\title{Reentrant vortex lattice transformation in four-fold
symmetric superconductors}

\author{N. Nakai, P. Miranovi\' c, M. Ichioka and  K. Machida}
\affiliation{Department of Physics, Okayama University, 700-8530
Okayama,
Japan}

\date{\today}

\begin{abstract}
The physics behind the rhombic$\rightarrow$square$\rightarrow$rhombic
flux line lattice transformation in increasing fields is clarified
on the basis of Eilenberger theory.
We demonstrate
that this reentrance observed in
LuNi$_2$B$_2$C is due to intrinsic competition
between superconducting gap and Fermi surface anisotropies.
The calculations reproduce
not only it but also predict yet not found lock-in
transition to a square lattice with different orientation in higher
field.
In view of physical origin given, this sequence of transitions is
rather
generic to occur in four-fold symmetric superconductors.
\end{abstract}

\maketitle

The morphology of equilibrium flux line lattice (FLL) in type-II
superconductors, its symmetry and orientation relative to the
crystallographic axes, is determined by microscopic electronic
properties, particularly by Fermi surface topology and superconducting
pair symmetry.
In general, the shape and orientation of FLL cell will change with
field direction trying
to adjust the electronic anisotropy of underlying crystal
\cite{weber}.
In spite of a long research history which dates back to the
pioneering small angle neutron scattering experiment by
Cribier {\it et al.} \cite{cribier} on Nb, problems
associated with FLL are still lively discussed even on seemingly
simplest elemental metal Nb \cite{forgan}.
FLL symmetry transformation under varying
applied field is one of the topics which attracted much of the
attention of vortex
physics community recently.
The effect became known in the early 70's when the low
field
rhombic-to-square FLL transition has been observed first in PbTl
\cite{obst}.
Renewed interest in this phenomena came, after similar transformations
has
been detected in a number of superconductors:
(Re)Ni$_2$B$_2$C (Re=Lu,Y,Er,Tm) \cite{er_square,koshelev,tm_square},
V$_3$Si \cite{mona} and
high-$T_c$ cuprate La$_{1.83}$Sr$_{0.17}$CuO$_4$ \cite{gilardi}.
All of them appear to be four-fold
symmetric crystals: cubic or tetragonal.

Recently, Eskildsen {\it et al.} \cite{eskildsen} discovered
a remarkable reentrant transition of the
rhombic FLL symmetry for ${\bm H} \parallel {\bm c}$
in LuNi$_2$B$_2$C ($H_{c2,\; 0} \sim 9\;$T).
Upon increasing $H$, the rhombic lattice changes into square one and
then
backs again at a higher field.
With increasing temperature the rhombus-square boundary bends away and
never crosses $H_{c2}$ line.
 Combining its result with other experiments
such as STM \cite{koshelev,sakata}, $\mu$SR \cite{ohishi} and
Bitter decoration \cite{vinikov}, square FLL region is confined in
a small area just above $H_{c1}$ and up to $\sim$ 2-3 T on $(H,T)$
plane.
High field square-to-rhombus transition is
detected in TmNi$_2$B$_2$C \cite{tm_square} as well, though this is a
magnetic
member of borocarbide family.
It is quite interesting to remember Nb case. The FLL in this cubic
system for
$\bm H \parallel (001)$ exhibits square-to-rhombic transition as $T$
increases
\cite{christen}. Although the definite phase diagram is not established
yet,
temperature dependence of FLL symmetry alone suggests an isolated region
of
stable square lattice in $H$ vs. $T$.

It was realized early \cite{takanaka}
that a certain four-fold anisotropy in plane perpendicular to
the applied field,
such as the Fermi velocity $v_F$, can drive
low field rhombus-to-square transformation. In the in-plane
anisotropy modeled by
$v_F(\theta)=v_F({\pi\over 8})(1+\beta\cos{4\theta})$ ($\theta$ polar
angle),
square FLL $\Box_v$
with nearest neighbors oriented along the velocity minimum will be stabilized
in low fields if anisotropy degree $\beta$ is large enough.
The same is true for the four-fold gap anisotropy alone
$|\Delta(\bm r,\theta)|^2
=|\Delta(\bm r,{\pi\over 8})|^2(1-\alpha\cos{4\theta})$, when square FLL
$\Box_g$ (nearest neighbors along gap minimum) tends to be stabilized.
These are schematically shown in Fig.1.
The nontrivial question is why  square FLL is unstable
at high fields and what the actual phase diagram looks like?
Contrary to the answer given by Gurevich and Kogan
\cite{kogan_gurevich} who consider
it due to thermal fluctuation near $H_{c2}$, here we investigate two
indispensable anisotropic effects
on the same footing. The interplay of gap and Fermi surface
anisotropy indeed can give rise to the reentrant FLL transition
and further square lock-in transition in a higher field.

There is firm evidence in LuNi$_2$B$_2$C and YNi$_2$B$_2$C,
which we regard as essentially the same systems, to show
the existence of both anisotropies. As for the gap anisotropy,
various bulk measurements, including the $H$-dependent
linear specific heat coefficient $\gamma(H)$ (Volovik effect)
\cite{nohara},
the thermal conductivity \cite{boaknin},
and angle-integrate photo-emission \cite{yokoya},
all show a substantial gap anisotropy.
More recently the angle-resolved thermal conductivity \cite{matsuda}
under $H$ exhibits that the gap is vanishingly smaller in $(100)$ than
in $(110)$. This is consistent with the spatial extension of the
zero-bias
peak observed by STM \cite{nishida}.
As for the Fermi velocity anisotropy,
when interpreted through simple four-fold harmonic variation,
band calculation \cite{dugdale}
told us  that the Fermi velocity is larger
in $(100)$ than in $(110)$, or the angle resolved
density
of states (DOS) $N(\theta)\sim 1/v_F(\theta)$ is smaller along $(100)$
compared to
$(110)$. Note that the sense of two anisotropies mentioned is what we
would
naively expected because the larger gap
$\Delta(\theta=0)$ should develop in
larger $N(\theta=0)$.
The built-in tendency to stabilize two different orientation
$\Box_v$ and $\Box_g$ of square lattice ultimately leads to the rich
vortex phase diagram, including the reentrant as we will see shortly.

Apart from the limiting cases:
London model at $H\ll H_{c2}(T)$
and Ginzburg-Landau (GL) model at $T\approx T_c$,
there is no handy and
convenient approximate scheme to
describe microscopic vortex properties deep in $(H,T)$ plane.
Therefore we resort to quasi-classical Eilenberger equations
\cite{eilenberger} valid for $k_F\xi_0\gg 1$ ($k_F$: the Fermi wave
number and $\xi_0$: the coherence length) a condition met in
most of superconductors. Eilenberger equations read as ($\hbar=1$)
\begin{eqnarray}
\left(2\omega+\bm v_F(\theta)\cdot\bm \Pi\right)
f(\omega,\bm r,\theta)=2\Delta(\bm r,\theta)
g(\omega,\bm r,\theta),
\end{eqnarray}
\begin{eqnarray}
\left(2\omega-\bm v_F(\theta)\cdot\bm \Pi^*\right)
f^\dagger(\omega,\bm r,\theta)=2\Delta^*(\bm r,\theta)
g(\omega,\bm r,\theta).
\end{eqnarray}
Here $\bm \Pi=\bm\nabla+(2\pi i/\Phi_0)\bm A$ is gauge invariant
gradient,
$\bm A$ is vector-potential and $\Phi_0$ is flux quantum.
$\omega=\pi T(2n+1)$ with integer $n$ is Matsubara frequency.
Normalization condition for Green's function $g^2+ff^\dagger=1$.
The pairing interaction is assumed separable $V(\theta,\theta')=
\overline{V}\phi(\theta)\phi(\theta')$ so that gap function
is $\Delta(\bm r,\theta)=\Psi(\bm r)\phi(\theta)$.
We consider two-dimensional case with cylindrical Fermi surface.
Four-fold models for Fermi velocity
$v_F(\theta)=\overline{v}_F(1+\beta\cos{4\theta})/\sqrt{1-\beta^2}$ and
gap anisotropy $\phi(\theta)^2=\phi({\pi\over 8})^2(1-\alpha\cos{4\theta})$
have been adopted.
Here $\theta$ is polar angle relative to $(100)$ axis.
Constant $\phi({\pi\over 8})^2=1/(1+(1-\sqrt{1-\beta^2})\alpha/\beta)$ is
chosen to
assure the same $T_c$ and DOS $N_0$ for any
value of anisotropy parameters $\alpha$ and $\beta$.
We are most interested in case when $\alpha$ and $\beta$
are of the same sign. Then positions
of gap and velocity minima are $45^\circ$ rotated to each other (see Fig.1)
and we have a competing effect.
The self-consistent equations for the gap function $\Psi(\bm r)$ and
vector-potential $\bm A$ are
\begin{figure}[t]
\includegraphics[height=4cm,keepaspectratio]{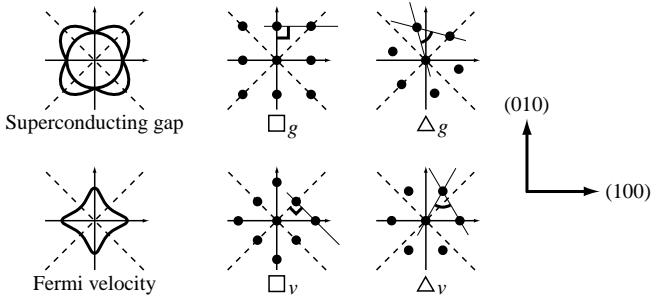}
\caption{Rhombus and square FLL cell orientations relative to the crystal. The
vortex center is shown by $\bullet$. $\alpha >0$, $\beta >0$.}
\label{nakai1}
\end{figure}
\begin{equation}
\Psi(\bm r)\ln\dfrac{T_c}{T}=2\pi T
\sum\limits_{\omega>0}\left[
\dfrac{\Psi(\bm r)}{\omega}-\left<
\dfrac{\phi(\theta) f}{v(\theta)}
\right>
\right],
\label{selfgap}
\end{equation}
\begin{equation}
\bm\nabla\times\bm\nabla\times\bm A=
-\dfrac{16\pi^3}{\Phi_0}N_0T\overline{v}_F
\sum\limits_{\omega>0}{\rm Im\;}\left<g\bm u\right>.
\end{equation}
Here, $v(\theta)=v_{F}(\theta)/\overline{v}_F$ and $\bm u=
(\cos\theta,\sin\theta)$ is
unit vector along the Fermi velocity $\bm v_F=v_F(\theta)\bm u$.
For average over Fermi surface,
$\left<\ldots\right>=(1/2\pi)\int\ldots d\theta$. Extra factor $1/v(\theta)$
in averages came from angle resolved DOS
$N(\theta)=N_0/v(\theta)$
on Fermi surface.

We have performed extensive numerical computations by the so-called
explosion
method (see Ref. \onlinecite{ichioka} for details) for various values
of anisotropy parameters $\alpha$ and $\beta$ in a high GL parameter case
$\kappa$=100.
The self-consistent
solution yields a complete set of the physical quantities: the spatial
profiles of the order parameter $\Psi(\bm r)$ and the magnetic field
$H(\bm r)$, and the local density of states around a vortex core.
The free energy density is given by
\begin{eqnarray}
F=\dfrac{\overline{H^2(\bm r)}}{8\pi}-2\pi TN_0\sum\limits_{\omega>0}
\left<\overline{
\dfrac{1-g}{1+g}\dfrac{\phi(\theta)(\Psi^*f+\Psi f^\dagger)}{2
v(\theta)}}
\right>.
\end{eqnarray}
Here, $\overline{a}=(B/\Phi_0)\int_{cell} a\;d\bm r$.
Free energy should be minimized with
respect to the FLL symmetry and its orientation relative to the
crystallographic axes. Numerics  is backed up by analytical
calculations.
Namely we also solve these analytically at the two limiting cases;
(1) near $H_{c2}(T)$ and (2) $H\ll H_{c2}(T)$ to gain physical insights.
For analytical results we considered FLL cell shaped as rhombus with
apex
angle in interval $[60^\circ,90^\circ]$. Two different orientations
of rhombus cell are compared: rhombus diagonals along gap minimum
(velocity maximum) and along velocity minimum (gap maximum)
(see Fig. \ref{nakai1}).
Since the numerical computation is very demanding and time
consuming, we limit ourselves to the four configurations
(apex angles $60^\circ$ and $90^\circ$ for each of two orientations)
as shown in Fig.1.

\begin{figure}[t]
\includegraphics[height=4cm,keepaspectratio]{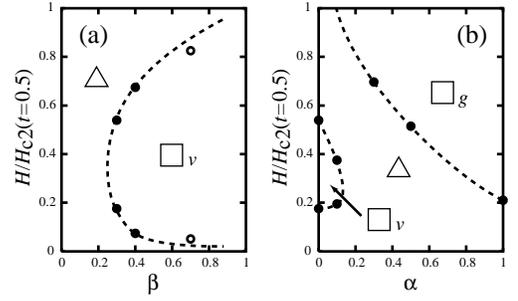}
\caption{FLL phase diagram at \mbox{$T/T_c=0.5$}.
a) $\beta$-dependence  ($\alpha$ = 0), $\circ$ shows where $\Box_v$ is stable.
b) $\alpha$-dependence
($\beta$ = 0.3). Broken lines are guide for the eye.}
\label{nakai2}
\end{figure}
In Fig. \ref{nakai2}(a)
we show the phase diagram for $t=T/T_c=0.5$ and $\alpha=0$, where the
square lattice $\Box_v$ becomes stable if the Fermi velocity anisotropy
$\beta$ exceeds a certain value. It shows also that the triangular lattice is
always stabilized at lower and higher $H$. Thus even without the gap
anisotropy ($\alpha$ = 0) the reentrant transformation can be induced.
This can be understood by looking at the self-consistency Eq.
(\ref{selfgap}).
It is seen that factor $1/v(\theta)$ plays the role of gap anisotropy
even in isotropic pairing case ($\phi(\theta)=1$).

The gap anisotropy further induces a rich variety of the
phase diagram. In Fig. \ref{nakai2}(b)
we show it when the moderate velocity anisotropy
$\beta=0.3$ is taken. We see that at the large $\alpha$ cases
the triangular lattice directly changes into the square lattice $\Box_g$.
This is a
similar situation to the $d$-wave case with the isotropic Fermi velocity
\cite{ichioka}. The preferred orientation perfectly coincides with the
nodal direction, namely, nearest neighbors of FLL
are along (100) (along (110)) in $d_{xy}$ ($d_{x^2-y^2}$) pairing.
As $\alpha$ decreases, this $\Box_g$ lattice region shrinks. At lower
fields yet another transformation from the triangular to square
$\Box_v$,
rotated by $45^\circ$ relative to $\Box_g$, emerges. The most
complicated case is at moderate values of $\alpha$ and $\beta$, leading to
successive lattice transformation $\triangle \rightarrow \Box_v
\rightarrow\triangle\rightarrow\Box_g$ as $H$ increases. In a weaker
$\alpha$
case the last lock-in transition is absent.

\begin{figure}[t]
\includegraphics[height=4cm,keepaspectratio]{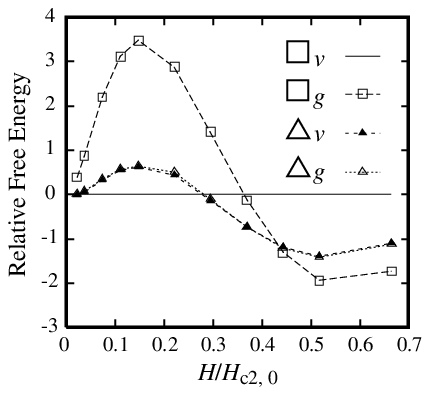}
\caption{Field dependence of free energy for $\triangle_v$, $\triangle_g$
and $\Box_g$ relative to $\Box_v$. $T/T_c=0.25$,
$\alpha=\beta=0.3$.
Free energies of $\triangle_{v}$ and $\triangle_{g}$
are almost identical.}
\label{nakai3}
\end{figure}

\begin{figure}
\includegraphics[height=5cm,keepaspectratio]{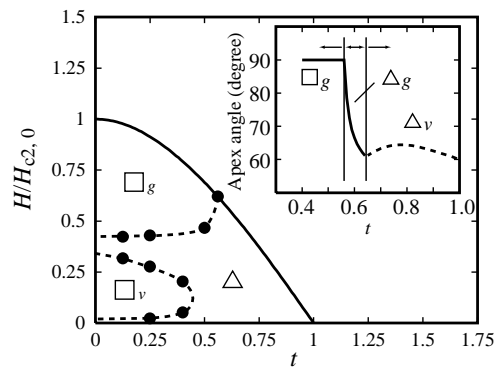}
\caption{Phase diagram of FLL for $\alpha=\beta=0.3$.
In inset, apex angle variation along $H_{c2}$ line.
The solid (broken) line is for orientation ``v'' (``g''),
see Fig. \ref{nakai1}.}
\label{nakai4}
\end{figure}

As a typical example we investigate the case $\alpha=\beta=0.3$
in details at lower temperature.
In Fig. \ref{nakai3} the four free energy curves are shown as a function
of $h=H/H_{c2,\; 0}$ at $t=0.25$. As $h$ increases the above mentioned
successive
transformations are clearly illustrated as several crossings.
Figure \ref{nakai4} displays
the resulting phase diagram in ($H$, $T$) plane. It is seen from this
that
(1) the $\Box_v$ region is confined to lower $H$ and $T$; (2) its
boundary
bends away from $H_{c2}$; (3) the high field region is occupied by the
$\Box_g$ lattice; (4) along $H_{c2}$ line this terminates at $t=0.56$,
below which the rhombic lattice becomes stable.

Let us discuss the physical origin of this intricate phase diagram in
connection
with the observation in borocarbides. Basically, in order to induce
narrowly limited $\Box_v$ region at low $H$ and $T$ we need the
competing
effects, each coming from the gap and velocity anisotropies.
The gap anisotropy with $\alpha>0$ prefers the lattice $\Box_g$ with
nearest
neighbors along (100) direction, while the velocity anisotropy with
$\beta>0$ tends to favor $45^\circ$ rotated square lattice $\Box_v$.
Thus these anisotropies compete each other.
The $\alpha\beta<0$ case does not cause such a frustration in FLL.

The interplay of two anisotropies can be understood
by looking at the free energy density at lower
$H$:
\begin{eqnarray}
F=\dfrac{B^2}{8\pi}\sum\limits_{\bm q}
\dfrac{1}{1+\lambda^2 q^2+\lambda^4(t_{xxyy}q^4+dq_x^2q_y^2)}
\end{eqnarray}
with $\bm q$ is the reciprocal vector of FLL and
\begin{eqnarray}
t_{ijlm}=\dfrac{4\pi^4N_0\overline{v}_F^4T\Psi_0^2}{\Phi_0^2}
\sum\limits_{\omega>0}\left<
\dfrac{\phi^2(\theta)v^3(\theta)u_iu_ju_lu_m}{
(\omega^2+\phi^2(\theta)\Psi_0)^{5/2}}
\right>,
\end{eqnarray}
$\lambda$ is penetration depth, and $\Psi_0(T)$ is zero field gap.
The extra four-fold anisotropy due to non-local
correction
appears through the parameter $d=t_{xxxx}+t_{yyyy}-6t_{xxyy}$ in the
above. At lower $T$, $t_{ijlm}$ contains factor
$v^3(\theta)/\phi^3(\theta)$
which strengthens the tendency towards the velocity anisotropy ($\beta$
is effectively increased by $\alpha$ when $\alpha\beta>0$). Near $T_c$ this
factor becomes $\phi^2(\theta)v^3(\theta)$ which weakens the combined
anisotropy effect by canceling each other. Therefore the two
anisotropies plays
different role, depending on $T$, giving rise to the bent
transition curve in the ($H$, $T$) plane.
For a fixed temperature
as $H$ increases from
$H_{c1}$, FLL starts with the regular triangle lattice because the
electromagnetic interaction between far apart vortices is isotropic,
yielding
the closed packed hexagonal symmetry just above $H_{c1}$.
As vortices approach each other, proliferating anisotropy in current
distribution will squeeze rhombus FLL cell toward square shape.
Being dominant at low $T$ region, anisotropy of Fermi velocity will
stabilize square lattice $\Box_v$ as soon as vortices come close to
each other with increasing $H$. The observed orientation of $\Box_v$
with nearest neighbors along (110) is indeed expected in LuNi$_2$B$_2$C
since the band
structure calculation indicates
that Fermi velocity is larger in (100) than (110)
when mapped into our four-fold model \cite{dugdale}. 
The same result is
obtained within nonlocal London model with Fermi velocity anisotropy
alone
\cite{kogan}. Upon further increasing $H$, condensation energy gradually
takes
over the major role
in determining the interaction between vortices.
Physically
it is due to the kinetic energy
cost of quasi-particles localized around core
(see Ref. \onlinecite{ichioka} for details).
In high fields, at least in high-$\kappa$ superconductors,
the anisotropy in vortex-vortex interaction is exclusively
due to the vortex core anisotropy giving rise to the transition
from the low field $\Box_v$ to high field $\Box_g$ via
intermediate rhombic lattice ($\Box_v\rightarrow\triangle
\rightarrow\Box_g$).

Limit of $H\approx H_{c2}(T)$ allows the
analytical solution for free energy
\begin{eqnarray}
F=
\dfrac{B^2}{8\pi}-\dfrac{1}{8\pi}
\dfrac{(B-H_{c2})^2}{{\cal F}+1},
\end{eqnarray}
where
\begin{eqnarray}
{\cal F}=\dfrac{2\pi^2N_0T}{\overline{h}_s^2}
\sum\limits_{\omega>0}\left<
\overline{
\dfrac{\phi(\theta)}{v(\theta)}ff^\dagger(\Psi f^\dagger+
\Psi f)}
\right>-\dfrac{\overline{h_s^2}}{\overline{h}_s^2}.
\end{eqnarray}
Here, $f$, $f^\dagger$ and $\Psi$ are solutions of the linearized
Eilenberger
equation, $h_s$ is magnetic field due to supercurrents,
$B=H+\overline{h}_s$
is magnetic induction.
In the isotropic case $\alpha=\beta=0$ and near $T_c$ it reduces to the
standard Abrikosov expression ${\cal F}=(2\kappa^2-1)\beta_A$.
This expression for free energy is valid all along the $H_{c2}(T)$ line.
The result is illustrated in Fig. 4 as the inset showing that the apex
angle
of the rhombic lattice continuous changes about the regular triangular
lattice
with $60^\circ$. The diagonal of the rhombic lattice rotated by
$45^\circ$
at $t=0.64$ from $(100)$ to $(110)$, {\it i.e.} from $\triangle_v$
to $\triangle_g$. At a higher field $\triangle_g$
lattice is locked in $\Box_g$. This lock-in point $t=0.56$ is rather nicely
connected
to the points determined numerically as shown in the main panel in Fig.
4.
This implies that our numerical results, which examine the limited
number
of lattice configurations (4 types), yield a reasonably phase diagram
even
taking into account general rhombic lattices.

The present calculation does not aim to quantitatively reproduce
the actual FLL phase diagram in LuNi$_2$B$_2$C,
but to physically understand its possible lattice transformation.
Note that so far the observed square lattices in LuNi$_2$B$_2$C
by several methods are all $\Box_v$, not $\Box_g$. The $\Box_g$ phase
can occur at further high $H$ if the gap anisotropy is large enough. Since
this is the case
for LuNi$_2$B$_2$C as mentioned before, there is a good chance to observe
it.
According to various thermodynamic and transport experiments the gap
anisotropy
is extremely large, the recent directional dependent measurements shows
the node situated at (100) direction. This unambiguously tells us that
LuNi$_2$B$_2$C belongs to the most interesting competing case
($\alpha>0$ and $\beta>0$) and also gratifying enough that it belongs to
the naively expected case where maximum energy gap is oriented to the
maximum DOS direction. So far the investigated field for the
vortex structure is limited below 4T (compare $H_{c2,\; 0}\sim 9$ T).

As for the other materials where the square lattice is found such as
Nb \cite{christen}, V$_3$Si \cite{mona} and Sr$_2$RuO$_4$ \cite{riseman}
we can expect the reentrant transition from the square to rhombic
lattice and further lock-in transition to the square lattice with
different orientation if the gap anisotropy is strong enough. The recent
finding of rhombus-to-square transition at lower $H$ in
La$_{1.83}$Sr$_{0.17}$CuO$_4$ \cite{gilardi} deserves a special attention
because in spite of $d_{x^2-y^2}$ symmetry they discovered $\Box_v$, not
$\Box_g$. This nontrivial observation is indeed expected by our
calculation
\cite{nakai}.

In conclusion we have shown that the vortex lattice morphology is deeply
connected to the underlying microscopic electronic structure.
Specifically
it is seen that the reentrant transition from the square to rhombic
lattice in LuNi$_2$B$_2$C can be well understood as arising from the
two competing superconducting gap and Fermi surface anisotropies both of
which are documented
to exist experimentally. It is demonstrated by solving the
quasi-classical
Eilenberger equations numerically and analytically. We have shown
another
yet not found FLL transformation to the differently oriented square
lattice in a higher field. In view of physical origin given, this kind
of re-entrance and high field square lattice are rather generic to occur.
Thus we
expect a similar successive transition in type-II superconductors with
four-fold symmetry at least with large $\kappa$.

We thank Y. Matsuda, N. Nishida, K. Takanaka and E. M. Forgan
for valuable information.


\begin{thebibliography}{99}
\bibitem{weber} For a review see {\it Anisotropy effects in
superconductors},
ed by H. Weber (Plenum, New York, 1977).
\bibitem{cribier}D. Cribier {\it et al.}, Phys. Lett. {\bf 9}, 106
(1964).
\bibitem{forgan} E. M. Forgan {\it et al.}, Phys. Rev. Lett. {\bf 88},
167003
(2002).
\bibitem{obst} B. Obst, Phys. Stat. Sol. {\bf 45}, 453 (1971).
\bibitem{er_square} U. Yaron {\it et al.}, Nature {\bf 382}, 236 (1996).
\bibitem{koshelev} Y. De Wilde {\it et al.}, Phys. Rev. Lett. {\bf 78},
4273 (1997).
\bibitem{tm_square} M. R. Eskildsen {\it et al.}, Nature {\bf 393},
242 (1998).
\bibitem{mona} M. Yethiraj {\it et al.}, Phys. Rev. Lett. {\bf 82},
5112 (1999).
\bibitem{gilardi} R. Gilardi {\it et al.}, cond-mat/0204278.
\bibitem{eskildsen} M. R. Eskildsen {\it et al.}, Phys. Rev. Lett. {\bf
86},
5148 (2001).
\bibitem{sakata} H. Sakata {\it et al.}, Phys. Rev. Lett. {\bf 84},
1583 (2000).
\bibitem{ohishi} K. Ohishi {\it et al.}, cond-mat/0201038.
\bibitem{vinikov} L. Ya. Vinnikov {\it et al.}, Phys. Rev. B
{\bf 64}, 024504 and 220508(R) (2001).
\bibitem{christen} D. K. Christen {\it et al.}, Phys. Rev. B {\bf 21},
102 (1980).
\bibitem{takanaka} K. Takanaka, Prog. Theor. Phys. {\bf 46},
1301 (1971).
\bibitem{kogan_gurevich} A. Gurevich and V. G. Kogan, Phys. Rev. Lett.
{\bf 87}, 177009 (2001).
\bibitem{nohara}M. Nohara {\it et al.}, J. Phys. Soc. Jpn. {\bf 66},
1888
(1997); K. Izawa {\it et al.}, Phys. Rev. Lett. {\bf 86}, 1327
(2001).
\bibitem{boaknin} E. Boaknin {\it et al.}, Phys. Rev. Lett.
{\bf 87}, 237001 (2001).
\bibitem{yokoya} T. Yokoya {\it et al.}, Phys. Rev. Lett. {\bf 85},
4952 (2000).
\bibitem{matsuda} Y. Matsuda, private communication.
\bibitem{nishida} N. Nishida, private communication.
\bibitem{dugdale} See for example, S. B. Dugdale {\it et al.},
Phys. Rev. Lett. {\bf 83}, 4824 (1999).
\bibitem{eilenberger} G. Eilenberger, Z. Physik {\bf 214}, 195 (1968).
\bibitem{ichioka} M. Ichioka {\it et al.}, Phys. Rev. B {\bf 59},
8902 (1999); M. Ichioka {\it et al.}, J. Phys. Soc. Jpn.
{\bf 66}, 3928 (1999).
\bibitem{kogan} V. G. Kogan {\it et al.}, Phys. Rev. B {\bf 55},
R8693 (1997).
\bibitem{riseman} T. M. Riseman {\it et al.}, Nature {\bf 396},
242 (1998).
\bibitem{nakai} N. Nakai {\it et al.}, in preparation.
\end{thebibliography}
\end{document}